\documentclass[11pt]{article}
\usepackage{txfonts}
\usepackage{cospar}
\usepackage[sectionbib]{natbib}
\renewcommand{\bibsection}{\section*{REFERENCES}}
\renewcommand{\bibhang}{\setlength{8mm}}
\renewcommand{\bibsep}{\setlength{0mm}}
\usepackage[dvips]{epsfig}
\usepackage{graphicx}
\usepackage{array}
\def\ts{\thinspace}					    
\def\cl{\centerline}					    
\def\R{{\it RXTE}}					    
\def\H{HEXTE}						    
\def\Z{\sf Z}						    
\def\hxt{hard X--ray tails}				    
\def\S{Sco{\ts}X-1}
\def\C{Cyg{\ts}X-2}
\def\9{GX{\ts}349$+$2}
\def\5{GX{\ts}5$-$1}
\def\7{GX{\ts}17$+$2}
\def\0{GX{\ts}340$+$0}
\def\M{$\dot{\hbox{M}}$}
\def\gtsim{\lower.5ex\hbox{$\; \buildrel > \over \sim \;$}}
\def\ltsim{\lower.5ex\hbox{$\; \buildrel < \over \sim \;$}}

\title{LONG TERM STUDIES OF {\Z} SOURCES \\ 
       WITH HEXTE/RXTE}

\author{F. D'Amico\address{INPE, Av. dos Astronautas 1758, 12227--010 S. J. dos Campos - SP, Brazil},
        W. A. Heindl\address{UCSD/CASS, 9500 Gilman Dr., La Jolla, CA 92093-0424, USA}, 
        and R. E. Rothschild$^2$}

\begin{document}

\maketitle

\begin{abstract}
We have analyzed the long pointed observations of the {\Z}
sources in the {\it Rossi X-Ray Timing Explorer (RXTE)} public archive to study
the high energy emission in those sources. Our analysis is concentrated on
the High Energy X--Ray Timing Experiment (HEXTE) waveband, since we are
primarily interested in studying the hard X--ray (i.e., $E >
20${\ts}keV) production in those sources. We give here the preliminary
results of this ongoing study. We have found no {\hxt} (besides {\S})
in our database from any of the {\Z} sources, i.e., {\9} ($< \, 7.9
\times 10^{-5}${\ts}photons{\ts}cm$^{-2}${\ts}s$^{-1}$, $3\sigma$,
50--150{\ts}keV), {\C} ($< \, 8.4 \times
10^{-5}${\ts}photons{\ts}cm$^{-2}${\ts}s$^{-1}$, $3\sigma$,
50--150{\ts}keV), {\7} ($< \, 4.2 \times
10^{-5}${\ts}photons{\ts}cm$^{-2}${\ts}s$^{-1}$, $3\sigma$,
50--150{\ts}keV), {\5} ($< \, 2.1 \times
10^{-5}${\ts}photons{\ts}cm$^{-2}${\ts}s$^{-1}$, $3\sigma$,
50--150{\ts}keV), and {\0} ($< \, 6.0 \times
10^{-5}${\ts}photons{\ts}cm$^{-2}${\ts}s$^{-1}$ $3\sigma$,
50--150{\ts}keV).  From the point of view of {\H}/{\R} observations
shown here, the production of {\hxt} in {\Z} sources is a process
triggered when special conditions are fulfilled. One of these
conditions, as derived from our analysis, is a threshold of 
$\sim 4 \times 10^{36}${\ts}erg{\ts}s$^{-1}$ for the luminosity of the
source's thermal component.
\end{abstract}

\section{INTRODUCTION}

The class of {\Z} sources comprises 6 Galactic LMXBs (i.e., {\S},
{\C}, {\9}, {\5}, {\7} and {\0}) which are accreting mass via Roche
lobe overflow. The inclusion of Cir{\ts}X-1 in this list remains
questionable (see, e.g., Iaria et al., 2002), and recently LMC{\ts}X-2
was discovered as the first {\Z} source outside our Galaxy (Smale, Homan and
Kuulkers, 2003). Another feature of these sources is that the inferred neutron
star magnetic field strength is intermediate ($B \sim 10^9${\ts}G)
between the range for the atoll sources ($\sim${\ts}$10^8$) and
accreting pulsars ({\gtsim}{\ts}$10^{12}$). Also, all of the {\Z} sources
besides {\S}, {\C}, and of course the recently found LMC{\ts}X-2 are
near or in the Galactic plane. They are called {\Z} sources due to
the track that they display in a color--color diagram
(see, e.g., van der Klis, 1996).  The movement along the {\Z} occurs in a
continuous fashion, in the sense that a source {\it must} spend some
time in the normal branch (NB) if it is moving from the horizontal
branch (HB) to the flaring branch (FB), i.e., a jump from the HB to
the FB (or vice--versa) has never been observed. It is believed that
the movement in the {\Z} is due to variations in the mass accretion
rate ({\M}, see, e.g., van der Klis, 1996), which increases in the
sequence HB$\longrightarrow$NB$\longrightarrow$FB.

The literature shows some examples of the detection of {\hxt} (HXT) in
{\Z} sources. In this article we will present our definition of a
HXT. For {\S} such detections were reported first by Strickman and
Barret (2000), and then confirmed and deeply studied by us (D'Amico et
al., 2001a, 2001b). A preliminary report of a HXT detection in {\C} was
presented by Frontera et al. (1998). Analyzing {\it BeppoSAX} data, Di
Salvo et al. (2000) reported the detection of a HXT in {\7}, and also
in {\9} (Di Salvo et al., 2001). We present the preliminary
results of an ongoing study that uses {\it Rossi X-Ray Timing Explorer}
({\R})/High Energy X-Ray Timing Experiment ({\H}) data in searching 
for HXT in {\Z} sources. In the next sections we present our
selected database and data analysis, and then our results are
discussed. It is interesting to highlight that we found {\it no} HXT in
the {\H} selected database (apart from {\S}; see D'Amico et al.,
2001a, 2001b). We will present our interpretations about this result.

\section{INSTRUMENTATION AND DATA ANALYSIS}

We used data from HEXTE (Rothschild et al., 1998) to search for hard
X--ray emission of the {\Z} sources in the $\sim${\ts}20--220{\ts}keV
interval and data from the Proportional Counter Array (PCA, Jahoda et
al., 1996) to determine the position of the source in the {\Z}
diagram (although such positions are not shown here in this
article). We have chosen, from the public {\it RXTE} database those
subsets of data in which $\gtsim${\ts}2000{\ts}s of HEXTE total
on--source time was available, in order to achieve better
sensitivity. The number of selected observations for each source is
shown in Table 1.

As we said in the previous section, almost all of the {\Z} sources are near
or in the Galactic Plane. The background in this particular region of
the sky is known to vary as a function of the latitude $b$, from the
``standard'' soft X--ray region (i.e., up to $\sim${\ts}20{\ts}keV,
see Valinia and Marshall, 1998) up to $\gamma$--ray energies (see,
e.g., Boggs et al., 2000).  We took advantage of the {\H} design to
carefully manage this problem by measuring the background at the same
latitude of the source's location, a feature that is possible only
because {\H} has two clusters that measure the background in 4
different sky regions (dubbed {\it plus} and {\it minus}
regions). Source confusion is another issue, again carefully studied
for each particular source. When a source was present in one of the
two background sampled regions, their presence was easily identified
and the other region was used as the source's background estimate.

We analyzed our data using the {\sf FTOOLS} and {\sf XSPEC}. Since
we are most interested in the {\H} data, data from PCA was modeled (in
the spectral analysis) using available literature models for each
source. In order to construct an uniform database for all the 6 {\Z} sources,
{\H} data was modeled with a sum of two parts: a {\it soft} part (up
to $\sim${\ts}40{\ts}keV in most of the cases) which was modeled as
a simple thermal bremsstrahlung, and a {\it hard} part which was
modeled as a power law model. We refer the reader to D'Amico et
al. (2001a, 2001b) for a more detailed description of the data
analysis, since we use the same procedure applied in {\S} for the
remaining {\Z} sources.

\section{RESULTS}

As defined in our previous studies (see, e.g., D'Amico et al., 2001b)
we developed two criteria to determine the presence of a HXT in a
particular spectrum: 1) a signal to noise ratio (SNR) $\geq 5$ in the
75-220{\ts}keV range, and 2) a F-Test null significance for the
addition of the hard component at a level of $10^{-7}$ or less (we caution the
reader however, that F-test can, in certain conditions, drive to wrong 
conclusions: see, e.g., Protasov et al., 2002). We
claim that we detected a HXT {\bf only} when  both of these criteria
are fulfilled.

We found no HXT in any of the observations of the {\Z} sources, but
{\S}. In Table 1 a compilation of our results is presented. In order
to estimate luminosities, the distances to the sources must be
known. Distances to {\S} and {\C} were already given in literature
(Bradshaw, Fomalont and Geldzahler, 1999; Orosz and Kuulkers, 1999,
respectively). The distance to the remaining sources was adopted as
being 7.9{\ts}$\pm${\ts}0.3{\ts}kpc, the distance to the Galactic
Center (GC: McNamara et al., 2000) since that distance has been
generally used as an estimate for the distance to all {\Z} sources (see,
e.g., Bandyopadhyay et al., 1999).

The thermal component was easily detected by {\H} up to 50{\ts}keV.
Nevertheless, the detection level was {\it always} below $3\sigma$ in
the 50-75{\ts}keV band (except for one case for {\0}) for all the
{\Z}s but {\S} and {\5}. The {\5} source is {\it strongly}
contaminated by the presence of the nearby source GRS{\ts}1758$-$258,
discovered by the {\it GRANAT} satellite (Sunyaev et al., 1991) and
highly studied with {\it RXTE} (Smith, Heindl and Swank, 2002). At
this point of our study we can firmly say that we did not observe a
HXT in {\5}, but we can say nothing yet about the flux of {\5} in the
50-75{\ts}keV band.

\begin{table}
\setlength{\extrarowheight}{1ex}
\begin{center}
\cl{\bf Table 1. ~~HEXTE Results: Fluxes and Luminosities}

\medskip
\begin{tabular}{c c c c c c c c c c}
\hline
\hline
Source Name &   &  Number of observations &         Flux$^{a}$                 &  &  &   L$_{20-80}{}^b$  &  &  & L$_{20-50}{}^c$ \\
\hline
{\S}        &   &  28 & $^d$2.0{\ts}$\times${\ts}$10^{-3}$ &  &  &     6.7$^e$        &  &  & 4.5--9.0$^f$     \\
{\9}        &   &  10 &    $< \, 7.9$                      &  &  &
$< \, 6.8$            &  &  &  $< \, 3.10$           \\
{\C}        &   &  13 &    $< \, 8.4$                      &  &  &
$< \, 5.0$             &  &  & $< \, 2.10$           \\
{\7}        &   &  11 &    $< \, 4.2$                      &  &  &
$< \, 8.8$             &  &  & $< \, 2.01$           \\
{\5}        &   &  12 &    $< \, 2.1$                      &  &  &
$< \, 16.1$            &  &  & $< \, 2.97$           \\
{\0}        &   &  13 &    $< \, 6.0$                      &  &  &
$< \, 13.4$            &  &  & $< \, 2.08$           \\
\hline
\end{tabular}
\end{center}

\smallskip
\noindent
{\small{NOTE: bremsstrahlung $+$ power law model used}}

\smallskip
\noindent
{\small{$^a$ 3$\sigma$ upper limit power law flux in the 50-150{\ts}keV range, in units of 
$10^{-5}${\ts}photons{\ts}cm$^{-2}${\ts}s$^{-1}$ (otherwise noted)}}

\smallskip
\noindent
{\small{$^b$ 3$\sigma$ power law luminosity in the 20-80{\ts}keV range,
with power law index frozen at 2 (otherwise noted), in 10$^{35}${\ts}erg{\ts}s$^{-1}$}}

\smallskip
\noindent
{\small{$^c$ Luminosity of the thermal component in the 20-50{\ts}keV range, 
in 10$^{36}${\ts}erg{\ts}s$^{-1}$}}

\smallskip
\noindent
{\small{$^d$ averaged in the HXT detections of {\S}: see D'Amico et al. (2001b)}}

\smallskip
\noindent
{\small{$^e$ measured in {\S}: see D'Amico et al. (2001b)}}

\smallskip
\noindent
{\small{$^f$ when a HXT is detected in {\S}: see D'Amico et al. (2001b)}}
\label{tab1}
\end{table}

\section{DISCUSSION}

From Table 1, and comparing the observations of the remaining sources
with {\S}, it seems to be that our {\H} observations were sensitive
enough to detect HXT. In comparison with the {\it BeppoSAX}
results reported for HXT detections in {\C} (Frontera et al., 1998), {\7} (Di
Salvo et al., 2000), and {\9} (Di Salvo et al., 2001), our results can
be interpreted in terms of variability in the appearance of a HXT in
the source's spectrum, as we observed in {\S} on a 4 hour time-scale
(D'Amico et al., 2001b). 

All the sources were observed in all the branches, but {\9} (from which
the HB was never observed up to date). It's also interesting to note that
the {\S} results (D'Amico et al., 2001b) found {\it no} correlation
between the observation of a HXT and the position of the source in the {\Z}, contrary
to the {\it BeppoSAX} results for {\7} and {\9} (Di Salvo et al., 2000 and 2001, 
respectively).

As we pointed out in previous studies (D'Amico et al., 2001b, 2001c),
the chance of observing a HXT in {\S} is higher when the thermal
component (20-50{\ts}keV) of the spectrum is brighter. As we can see
in Table 1, our measured level for the thermal luminosities of the
{\Z} sources (besides {\S) is {\it always} below a level of
$\sim 4 \times 10^{36}$ erg{\ts}s$^{-1}$. While comparable values
were not given by the {\it BeppoSAX} results for {\C}, {\7}, and {\9}
(nor by the {\it CGRO} results for {\S}: see Strickman and Barret,
2000), it is possible to extrapolate {\it BeppoSAX} results for {\9}
in order to estimate the luminosity of the thermal component. We
estimate that the 20-50{\ts}keV luminosity {\9} luminosity measured by
{\it BeppoSAX} was greater than $\sim 5 \times 10^{36}$
erg{\ts}s$^{-1}$. It thus appear from the point of view of the {\H}
observations shown here, that the production of HXT in {\Z} sources is
a process triggered when special conditions are fulfilled. Our
observations show that one of this special conditions is the
brightness of the thermal component, being the HXT production
triggered when this luminosity is above the threshold of $\sim 4
\times 10^{36}$ erg{\ts}s$^{-1}$.

We still want to emphasize the case of {\S} as a peculiar {\Z}
source. Scorpius{\ts}X-1 remains the only {\Z} where a HXT was detected
more than once, and also by two different satellites. 

\section{CONCLUSIONS}

As we have shown here, our {\H} results on {\Z} sources show no
evidence for the presence of HXT in several long pointed
observations. We took advantage of the {\H} design to constrain the
contamination of the source's spectrum due to the variations of the
background with latitude in the Galactic ridge region, and also
carefully studied background and/or source contaminations by nearby
sources, as is the case for {\5} which is contaminated by the presence
of GRS{\ts}1758$-$258. We have shown that our observations were
sensitive enough to detect the presence of the HXT, in comparing with
our observations of HXT in {\S} and with other reported HXT detections
by {\it BeppoSAX}. Our interpretation for this is in terms of
variability of the sources. Such variations were observed in {\S} on a
4 hour time-scale. From the point of view of {\H} observations shown
here, and also from the {\it BeppoSAX} HXT detection in {\9} the
production of a HXT in a {\Z} source is a process triggered when, at
least, among other possible conditions, the brightness of the thermal
component is above a level of $\sim 4 \times 10^{36}$
erg{\ts}s$^{-1}$.

\section{ACKNOWLEDGMENTS}
This research has made use of data obtained through the High Energy
Astrophysics Science Archive Research Center Online Service, provided
by the NASA/Goddard Space Flight Center. This work was partially supported
by FAPESP/Brazil under grant 99/02352--2. F. D'Amico gratefully acknowledges
the hospitality of all CASS/UCSD members in the 1999--2001 period
when this project was conceived: thanks a lot!
This research was supported by NASA contract NAS5--30720.

\bigskip
\medskip
\medskip
\noindent
E-mail address of F. D'Amico: damico@das.inpe.br

\end{document}